\documentstyle[preprint,prc,aps,epsfig]{revtex}
\begin{document}
\draft
\tighten
\title{Density dependent hadron field theory for neutron stars with antikaon 
condensates}
\author{Sarmistha Banik and Debades Bandyopadhyay}
\address{Saha Institute of Nuclear Physics, 1/AF Bidhannagar, 
Kolkata 700 064, India}

\maketitle

\begin{abstract}
We investigate $K^-$ and $\bar K^0$ condensation in $\beta$-equilibrated
hyperonic matter within a density dependent hadron field 
theoretical model. In this model, baryon-baryon and (anti)kaon-baryon 
interactions are mediated by the exchange of mesons. Density dependent 
meson-baryon coupling constants are obtained from microscopic Dirac Brueckner
calculations using Groningen and Bonn A nucleon-nucleon potential. It is found 
that the threshold of antikaon condensation is not only sensitive to the 
equation of state but 
also to antikaon optical potential depth. Only for large values of antikaon 
optical potential depth, $K^-$ condensation sets in even in the presence of 
negatively charged hyperons. The threshold of $\bar K^0$
condensation is always reached after $K^-$ condensation. Antikaon 
condensation makes the equation of state softer thus resulting in smaller 
maximum mass stars compared with the case without any condensate.

{\noindent\it PACS}: 26.60.+c, 21.65.+f, 97.60.Jd, 95.30.Cq
\end{abstract}
\section{Introduction}
The composition and structure of neutron stars depend on the nature of strong 
interaction. Neutron star matter encompasses a wide range of densities, 
from the density of iron nucleus at the surface of the star to several times 
normal nuclear matter density in the core. Since the chemical potentials of
nucleons and leptons increase rapidly with density in the interior of neutron
stars, several novel phases of matter with large strangeness fraction such as, 
hyperonic matter, condensates of strange mesons and quark matter may 
appear there \cite{Gle97}. 

It was first demonstrated
by Kaplan and Nelson within a chiral $SU(3)_L \times SU(3)_R$ model that $K^-$ 
meson may undergo Bose-Einstein condensation in dense matter formed in heavy ion
collisions \cite{Kap}. In this model baryons directly couple with (anti)kaons.
The effective mass of antikaons decreases with increasing 
density because of the strongly attractive $K^-$-baryon interaction in dense
matter. Consequently, the in-medium energy of $K^-$ mesons in the zero-momentum
state also decreases with density. The $s$-wave $K^-$ condensation sets in 
when the energy of $K^-$ mesons equals to its chemical potential. Later, 
$K^-$ condensation in the core of neutron stars was studied  by other groups 
using chiral models \cite{Schm}. 

Also, Bose-Einstein 
condensation of $K^-$ mesons was investigated in the traditional meson exchange 
picture known as relativistic mean field (RMF) model \cite{Ell,Sch96,Kno}. 
Within the framework of RMF model, baryon-baryon and (anti)kaon-baryon 
interactions are treated in the same footing i.e. those are mediated by the 
exchange of mesons \cite{Sch99,Pal,Bani}. It was noted in all these 
calculations that the typical threshold density of $K^-$ condensation in 
nucleons-only neutron star matter was about 2-4$n_0$, where $n_0$ is normal 
nuclear
matter density. However, the threshold of antikaon condensation is sensitive 
to the antikaon optical potential and depends more strongly on the equation of
state. With further inclusion of hyperons, $K^-$ condensation was found to
occur at higher densities \cite{Ell,Sch96,Kno,Pal,Bani}. 
Recently, we have studied $\bar K^0$ condensation along with $K^-$ condensation
in neutron stars using a relativistic mean field model and its influence on 
the structure of neutron stars \cite{Pal,Bani}. The threshold density of 
$\bar K^0$ condensation is always higher than that of $K^-$ condensation.    
Antikaon condensate makes the equation of state (EoS) softer, thus resulting in
smaller maximum mass stars compared with the case without any condensate. 
Employing the EoS including both $K^-$ and $\bar K^0$ condensates, it was 
predicted 
\cite{Bani} that a stable sequence of superdense stars called the third 
family branch \cite{Ger} might exist beyond the neutron star branch. The 
compact stars in the third family branch have smaller radii than those of the 
neutron star branch \cite{Bani}.

Besides EoS and antikaon optical potential depth, the threshold density of 
antikaon condensation is very much sensitive to the behaviour of antikaon 
energy and electron chemical potential at high density. The role of 
nucleon-nucleon and (anti)kaon-nucleon correlation on antikaon condensation were
investigated by Pandharipande and collaborators \cite{Pan95,Pan01}. They found
that strong nucleon-nucleon and (anti)kaon-nucleon correlation raised the 
critical
density for antikaon condensation to higher densities and predicted antikaon
condensation might not be a possibility in neutron stars \cite{Pan01}. The 
electron chemical potential used in the above mentioned calculations was 
obtained from modern realistic nucleon-nucleon interactions \cite{Wir,Akm}.

In this work, we are interested to find out how many body
correlations which may be taken into account by density dependent meson-baryon 
couplings in a relativistic field theoretical model, affect the threshold of 
antikoan condensation in neutron star matter. 
There is a growing interest to derive a quantum hadron field theory 
from a microscopic approach to nuclear interactions. The motivation of such an 
approach is not only to retain the essential features of quantum hadrodynamics
(QHD) \cite{Ser}, but also to deal with the complicated many body 
dynamics of
strong interactions \cite{Bro92,Fuc95,Len95}. An appropriate and successful
microscopic approach to in medium nuclear interactions follows from 
Dirac Brueckner (DB) calculations. Various groups performed DB calculations 
with realistic nucleon-nucleon interactions and reproduced empirical saturation
properties of symmetric nuclear matter reasonably well 
\cite{Ana,Hor,Har,Mac,Bro90,Boe,Hub95,Len98}. Also nucleons-only neutron star
matter was calculated in Dirac Brueckner theory \cite{Mue,Eng,Hub98}.    
It is worth mentioning here that the bulk of 
the screening of nucleon-nucleon interaction in medium is taken into account by
the local baryon density dependent DB self energies \cite{Fuc95}. This makes
relativistic many body dynamics to be approximated by a density dependent 
relativistic hadron (DDRH) field theory \cite{Fuc95,Len95}. A covariant and 
thermodynamically consistent DDRH field theory is obtained by making 
interaction vertices as Lorentz scalar functionals of baryon field operators. 
In the mean field approximation this model reduces to the relativistic 
Hartree description with density dependent meson-nucleon
couplings. The density dependent meson-nucleon couplings are obtained from 
Dirac Brueckner self energies calculated with Bonn, Groningen and 
phenomenological density dependent potentials 
\cite{Had,Rin98,Typ,Kei,Hof01,Rin02}. The variational derivatives of vertices 
with respect to baryon fields give rise to rearrangement terms in baryon
field equations \cite{Len95}. Brockmann and Toki \cite{Bro92} first applied
DDRH model without rearrangement terms to study finite nuclei. Recently, DDRH
model with rearrangement terms has been exploited to investigate 
deformed nuclei \cite{Rin98}, hypernuclei \cite{Kei}, asymmetric nuclear matter
and exotic nuclei \cite{Hof01,Rin02} and neutron star properties \cite{Hof2}. 

In this paper, we investigate antikaon condensation in beta equilibrated 
hyperon matter relevant to neutron stars and its role on the composition and
structure of the compact stars in DDRH model. 
The paper is structured in the following way. In Sec. II, we describe DDRH 
model and different phases of matter. Parameters of the model are discussed in
Sec. III. Results of our calculation are explained in Sec. IV. And Sec. V 
provides a summary and conclusions. 

\section{Formalism}
Here, we discuss a phase transition from hadronic matter to antikaon condensed
phase in compact stars. The hadronic phase is described within the framework of
DDRH model. This phase is composed of all species of the baryon octet, 
electrons and muons. Therefore, the total Lagrangian density in the hadronic 
phase is written as ${\cal L} = {\cal L}_B + {\cal L}_l$. In DDRH model, 
baryon-baryon interaction is given by the  Lagrangian density (${\cal L}_B$)
\cite{Hof2},
\begin{eqnarray}
{\cal L}_B &=& \sum_B \bar\Psi_{B}\left(i\gamma_\mu{\partial^\mu} - m_B
+ g_{\sigma B} \sigma - g_{\omega B} \gamma_\mu \omega^\mu 
- \frac{1}{2} g_{\rho B} 
\gamma_\mu{\mbox{\boldmath $\tau$}}_B \cdot 
{\mbox{\boldmath $\rho$}}^\mu +\frac {1} {2} g_{\delta B}{\mbox{\boldmath 
$\tau$}}_B \cdot {\mbox{\boldmath $\delta$}} \right)\Psi_B\nonumber\\
&& + \frac{1}{2}\left( \partial_\mu \sigma\partial^\mu \sigma
- m_\sigma^2 \sigma^2\right) + \frac{1}{2}\left( \partial_\mu \delta\partial
^\mu \delta
- m_\delta^2 \delta^2\right)  
 -\frac{1}{4} \omega_{\mu\nu}\omega^{\mu\nu}\nonumber\\
&&+\frac{1}{2}m_\omega^2 \omega_\mu \omega^\mu
- \frac{1}{4}{\mbox {\boldmath $\rho$}}_{\mu\nu} \cdot
{\mbox {\boldmath $\rho$}}^{\mu\nu}
+ \frac{1}{2}m_\rho^2 {\mbox {\boldmath $\rho$}}_\mu \cdot
{\mbox {\boldmath $\rho$}}^\mu,
\end{eqnarray}
and
\begin{eqnarray}
{\cal L}_l &=& 
\sum_l \bar\psi_l\left(i\gamma_\mu {\partial^\mu} - m_l \right)\psi_l ~.
\end{eqnarray}

The field strength tensors for vector mesons are given by
\begin{eqnarray}
\omega^{\mu \nu}&=& \partial^ \mu \omega^ \nu-\partial^\nu \omega^ \mu 
\nonumber \\
\rho^{\mu \nu}&=& \partial^ \mu \rho^ \nu-\partial^\nu \rho^ \mu 
\end{eqnarray}

Here $\Psi_B$ denotes the isospin multiplets for baryons 
and the sum goes over baryon multiplets B = N,$\Lambda$,$\Sigma$,$\Xi$; 
$\psi_l$ ($l \equiv {e,\mu}$) is lepton spinor and ${\mbox{\boldmath 
$\tau_{B}$}}$ is isospin 
operator. The interactions among baryons are mediated by the exchange of 
$\sigma$, $\omega$ and $\rho$ mesons. In addition to these mesons the 
scalar-isovector meson $\delta$ is also included. And this is important for 
an asymmetric system. Though the structure of DDRH Lagrangian density closely
follows that of RMF model, there are important differences between those 
models. In RMF calculations with density independent meson-baryon coupling 
constants, non-linear self interaction terms for scalar and vector fields 
are inserted to account for higher order density dependent contributions. 
But this is not needed here as meson-baryon vertices $g_{\alpha B}$, where 
$\alpha$ denotes $\sigma$, $\omega$, $\rho$ and $\delta$ fields, are dependent 
on Lorentz scalar functionals of baryon 
field operators and adjusted to the Dirac-Brueckner-Hartree-Fock (DBHF) 
calculations \cite{Hof01,Hof2}. 

There are two choices for the density dependence of meson-baryon couplings. One 
is the scalar density dependence (SDD) and the other one is the vector density 
dependence (VDD) \cite{Hof01}. Here 
we consider meson-baryon couplings $g_{\alpha B}(\hat {\rho})$ to depend 
on vector density because it gives a more natural connection to the 
parameterization of DB vertices. For VDD case, the density operator 
$\hat {\rho}$ has the form, 
$\hat \rho$=$\sqrt{{\hat j}_\mu {\hat j}^{\mu}}$, where ${\hat j}_\mu = \bar 
\Psi \gamma_\mu \Psi$.
 
As vertices $g_{\alpha B}$s are Lorentz scalar functional of baryon field 
operators, the variation of $\cal L$ with respect to $\bar \Psi_B$ gives, 
\begin{eqnarray}
\frac {\delta \cal L} {\delta \bar \Psi_B}=\frac {\partial \cal L} {\partial 
\bar \Psi_B}+ \frac {\partial \cal L} {\partial {\hat \rho_B}}
\frac {\delta {\hat \rho_B}} {\delta \bar \Psi_B}
\end{eqnarray}
The rearrangement term 
$\Sigma^{\mu(r)} = \sum_B {\frac {\partial \cal L} {\partial {\hat \rho_B}}
\frac {\delta {\hat \rho_B}} {\delta \bar \Psi_B}}$ which originates from the
second term of Eq. (4),
naturally introduces additional contribution to vector self-energy 
\cite{Fuc95,Len95,Hof01,Hof2}. This is an important difference between RMF and 
DDRH theory. 
 
Here, we perform our calculation in the mean field approximation (MFA).
In this approximation vertex functionals 
are reduced to simpler forms using
Wick's theorem \cite{Hof01,Wi}. The operator $\hat \rho$ is 
replaced by ground state expectation value $\rho$ i.e. $<\hat \rho>=\rho$. 
Hence meson-baryon vertices become function of total baryon density in the 
hadronic phase, 
\begin{equation}
<g_{\alpha B}(\hat \rho)>=g_{\alpha B}(<\hat \rho>)=g_{\alpha B}(\rho).
\end{equation}
This is known as vector density dependence of vertices \cite{Fuc95,Hof01,Hof2}. 
In MFA adopted here, meson fields are replaced 
by their expectation values. Only the time-like components
of vector fields, and isospin 3-components of $\rho$ and $\delta$ 
fields have non-vanishing values in a uniform and static matter.
The mean meson fields are denoted by $\sigma$,  $\omega_0$,
$\rho_{03}$ and $\delta$. 
Therefore, the meson field equations in the hadronic (h) phase are given by
\begin{eqnarray}
m_\sigma^2\sigma &=& 
\sum_b g_{\sigma b} n_b^{h,s}~,\\
m_\omega^2\omega_0 &=& \sum_b g_{\omega b} n_b^h~,\\
m_\rho^2\rho_{03} &=&\frac{1}{2} \sum_b g_{\rho b} \tau_{3b} n_b^h~,\\
m_\delta^2\delta &=&\frac{1}{2} \sum_b g_{\delta b}\tau_{3 b} n_b^{h,s}~,
\end{eqnarray}
where $\tau_{3b}$ is the isospin projection of baryon b = n, p, $\Lambda$,
$\Sigma^-$, $\Sigma^0$, $\Sigma^+$, $\Xi^-$, $\Xi^0$
and 
scalar and vector densities of baryon b in the hadronic phase are 
\begin{eqnarray}
n^{h,s}_b &= & <\bar \psi_{b} \psi_b>= \frac{2J_b+1}{2\pi^2} \int_0^{k_{F_b}}
\frac{m_b^*}{(k^2 + m_b^{* 2})^{1/2}} k^2 \ dk \nonumber \\
&&=\frac{m_b^*}{2\pi^2}[k_{F_{b}}\sqrt{{k_{F_b}}^2+m_b^{*2}} - m_b^{*2}ln 
\frac {k_{F_{b}}+\sqrt{{k_{F_b}}^2+m_b^{*2}}}{m_b^*}]~,\\
n_b^h &=& <\bar \psi_b \gamma_0 \psi_b>=\frac{k_{F_{b}}^3}{3 \pi^2}~.
\end{eqnarray}
Here $J_b$ is the spin projection of baryon b.
The rearrangement self energy modifies the baryon field 
equation compared to the RMF case \cite{Ser},
\begin{equation}
[\gamma_{\mu}\left(i\partial^{\mu}-\Sigma^0_{b,h}\right)-\left(m_b
-\Sigma^s_{b,h}\right) ]\psi_{b}=0.
\end{equation}
Here $\psi_b$ is the Dirac spinor for baryon $b$. The total vector self energy 
for baryon b in the hadronic phase is 
\begin{equation}
\Sigma^0_{b,h}=\Sigma^{0(0)}_{b,h}+\Sigma^{0(r)}_h. 
\end{equation}
Now the usual vector self energy is
\begin{equation}
\Sigma^{0(0)}_{b,h}=g_{\omega b}\omega_0 +\frac {1} {2} g_{\rho b} \tau_{3b} 
\rho_{03}~.
\end{equation}
Also, the rearrangement term which is the second term in Eq. (4), simplifies to 
\cite{Hof01} 
\begin{equation}
\Sigma^{0(r)}_h=\sum_b[-\frac {\partial g_{\sigma b}} {\partial \rho_b} 
\sigma n^{h,s}_b + \frac { \partial g_{\omega b}}{\partial \rho_b} \omega_0 
n^h_b
+ \frac {1} {2}\frac { \partial g_{\rho b}}{\partial \rho_b }\tau_{3b} 
\rho_{03} n^h_b
-  \frac {1} {2}\frac { \partial g_{\delta  b}}{\partial \rho_b }\tau_{3b} 
\delta n^{h,s}_b  ]~,
\end{equation}
Similarly, the expression of scalar self energy for baryon b 
is given by 
\begin{equation}
\Sigma^s_{b,h}=g_{\sigma b}\sigma + \frac {1}{2} g_ {\delta b }\tau_{3b} 
\delta .
\end{equation}
One can immediately define effective baryon mass as $m_b^*=m_b-\Sigma^s_{b,h}$.
And this differs for members of isospin multiplets due to $\delta$ meson. 

To obtain the EoS (pressure versus energy density) of the pure hadronic phase, 
the equations of motion for mesons and baryons (Eqs. (6)-(9) and Eq. (12)) 
are solved self-consistently along with effective baryon mass
in the mean field approximation keeping 
into consideration other constraints such as baryon number conservation, 
charge neutrality and beta equilibrium. The system is
charge neutral and the condition of $\beta$-equilibrium is maintained. 
The charge neutrality condition is 
\begin{equation}
Q^h = \sum_b q_b n^h_b -n_e -n_\mu =0,
\end{equation}
where $q_b$ and $n_b^h$ are electric charge and the number density of baryon b 
in the pure hadronic phase, respectively and
$n_e$ and $n_\mu$ are number densities of electrons and muons respectively.
In compact star interior, chemical equilibrium is maintained through weak 
interactions such as 
$B_1 \longrightarrow B_2 + l+ \bar \nu_l$ and 
$B_2 +l \longrightarrow B_1 +\nu_l$ where $B_1$ and $B_2$ are baryons and 
l stands for leptons. 
Therefore the generic equation relating chemical potentials for the 
above mentioned generalised $\beta$-decay processes is
\begin{equation}
\mu_i = b_i \mu_n - q_i \mu_e ~,
\end{equation}
Here $b_i$ and $q_i$ are the baryon number and charge of ith baryon and 
$\mu_n$ and $\mu_e$ are the
chemical potentials of neutron and electron, respectively. The chemical 
potential of baryon b in the hadronic phase is expressed as
\begin{equation}
\mu_b=\sqrt{k_{F_b}^2+m_b^*2} +\Sigma^{0(0)}_{b,h}+\Sigma^{0(r)}_h~.
\end{equation}
It is noted that unlike RMF model, the rearrangement term appears in the 
expression of baryon chemical potential in DDRH model. 
In neutron stars, electrons are converted to muons by
$e^- \to \mu^- + \bar \nu_\mu + \nu_e$ when the electron chemical potential
becomes equal to the muon mass. Therefore, we have $\mu_e=\mu_{\mu}$ in 
compact stars. Equation (18) implies that there are two independent 
chemical potentials $\mu_n$ and $\mu_e$ corresponding to two conserved 
charges i.e. baryon number and electric charge.
The energy density ($\varepsilon^h$) is related to the pressure
($P^h$) in this phase through the Gibbs-Duhem relation 
\begin{equation}
P^h=\sum_i \mu_i n_i -\varepsilon^h~.
\end{equation}
Here $\mu_i$ and $n_i$ are chemical potential and number density for 
i-th species. The expression of energy density in the hadronic phase is
\cite{Hof2} 
\begin{eqnarray}
{\varepsilon^h}  &=& \frac{1}{2}m_\sigma^2 \sigma^2
+ \frac{1}{2} m_\omega^2 \omega_0^2+ \frac{1}{2} m_\rho^2 \rho_{03}^2+
 \frac{1}{2} m_\delta^2 \delta^2  \nonumber \\
&& + \sum_b \frac{2J_b+1}{2\pi^2} 
\int_0^{k_{F_b}} (k^2+m^{* 2}_b)^{1/2} k^2 \ dk
+ \sum_l \frac{1}{\pi^2} \int_0^{K_{F_l}} (k^2+m^2_l)^{1/2} k^2 \ dk. 
\end{eqnarray}
The rearrangement 
term does not contribute to the energy density explicitly, whereas it occurs in
the pressure through baryon chemical potential. 
It is the rearrangement term that accounts for the energy-momentum conservation 
and thermodynamic consistency of the system \cite{Len95}.

The pure antikaon (${\bar K}$) condensed phase is composed of baryons, 
leptons and antikaons
which are in chemical equilibrium under weak interactions and maintain local 
charge neutrality. The baryon-baryon interactions here are described by the 
Lagrangian density of DDRH model. It is worth mentioning here that 
the meson-baryon couplings depend on the total baryon density in this phase.
In this phase, baryons are embedded in 
antikaon condensates. Earlier it was noted that baryons in the pure 
hadronic and antikaon condensed phase behaved differently because of their 
dynamical nature \cite{Sch99,Pal,Bani}. It was attributed to different mean 
fields which baryons experienced in those pure phases. 
We adopt here a relativistic field
theoretical approach for the description of (anti)kaon-baryon interaction
\cite{Sch99,Bani}. In this model (anti)kaon-baryon interactions are 
mediated by $\sigma$, $\omega$, $\rho$ and $\delta$ meson. 
The Lagrangian density for (anti)kaon interaction
in the minimal coupling scheme is, 
\begin{equation}
{\cal L}_K = D^*_\mu{\bar K} D^\mu K - m_K^{* 2} {\bar K} K ~,
\end{equation}
where the covariant derivative
$D_\mu = \partial_\mu + ig_{\omega K}{\omega_\mu} + 
+ ig_{\rho K}
{\mbox{\boldmath $\tau$}}_K \cdot {\mbox{\boldmath $\rho$}}_\mu/2$. 
The isospin doublet for kaons
is denoted by $K\equiv (K^+, K^0)$ and that for antikaons is
$\bar K\equiv (K^-, \bar K^0)$. It is to 
be noted that the coupling constants of (anti)kaon-baryon interactions are 
considered to be density independent. The effective mass of (anti)kaons in this
minimal coupling scheme is given by 
\begin{equation}
m_K^* = m_K - g_{\sigma K} \sigma -\frac {1} {2} g_{\delta K}\tau_{3{\bar K}}
\delta ~,
\end{equation}
where $m_K$ is the bare kaon mass. Here also the effective mass of $K^-$ and 
$\bar K^0$ differ due to the inclusion of the scalar-isovector $\delta$ meson.
The dispersion relation representing the in-medium energies of
$\bar K\equiv (K^-, \bar K^0)$ for $s$-wave (${\bf k}=0$) condensation
is given by
\begin{equation}
\omega_{K^-,\: \bar K^0} = m_K^* - g_{\omega K} \omega_0   
\mp \frac{1}{2} g_{\rho K} \rho_{03} ~,
\end{equation}
where the isospin projection $\tau_{3 {\bar K}} =\mp 1$ for the mesons
$K^-$ ($-$ sign) and $\bar K^0$ (+ sign) are explicitly written in the
expression. For s-wave condensation, densities of antikaons are given by
\begin{equation}
n_{K^-,\: \bar K^0} = 2\left( \omega_{K^-, \bar K^0} + g_{\omega K} \omega_0 
\pm \frac{1}{2} g_{\rho K} \rho_{03} \right) {\bar K} K  
= 2m^*_K {\bar K} K  ~.
\end{equation}

In the mean field approximation, the meson field equations in the presence of 
antikaon condensates are given by
\begin{eqnarray}
m_\sigma^2\sigma &=& 
\sum_b g_{\sigma b} n_b^{{\bar K},s}
+ g_{\sigma K} \sum_{\bar K} n_{\bar K} ~,\\
m_\omega^2\omega_0 &=& \sum_b g_{\omega b} n_b^{\bar K}
- g_{\omega K} \sum_{\bar K} n_{\bar K} ~,\\
m_\rho^2\rho_{03} &=&\frac{1}{2} \sum_b g_{\rho b} \tau_{3b} n_b^{\bar K}
+\frac{1}{2} g_{\rho K} \sum_{\bar K} \tau_{3 {\bar K}} n_{\bar K}~,\\
m_\delta^2\delta &=&\frac{1}{2} \sum_b g_{\delta b}\tau_{3 b} n_b^{{\bar K},s}
+\frac{1}{2} g_{\delta K} \sum_{\bar K}\tau_{3 {\bar K}} n_{\bar K} ~,
\end{eqnarray}
where $n_b^{{\bar K},s}$ and $n_b^{\bar K}$ are scalar and vector density of 
baryon b in the antikaon condensed phase and have the same forms as in Eqs. (10)
and (11). The meson field equations here remain same 
in structure as RMF ones \cite{Sch99,Pal,Bani}, but the constant meson-baryon
couplings are replaced by their density dependent counterparts. 

The total energy density and pressure in the antikaon condensed phase 
are given by \cite{Pal,Bani}
\begin{eqnarray}
{\varepsilon^{\bar K}}  &=& \frac{1}{2}m_\sigma^2 \sigma^2
+ \frac{1}{2} m_\omega^2 \omega_0^2+ \frac{1}{2} m_\rho^2 \rho_{03}^2+
 \frac{1}{2} m_\delta^2 \delta^2  \nonumber \\
&& + \sum_b \frac{2J_b+1}{2\pi^2} 
\int_0^{k_{F_b}} (k^2+m^{* 2}_b)^{1/2} k^2 \ dk
+ \sum_l \frac{1}{\pi^2} \int_0^{K_{F_l}} (k^2+m^2_l)^{1/2} k^2 \ dk 
\nonumber \\
&& + m^*_K \left( n_{K^-} + n_{\bar K^0} \right)~ ,
\end{eqnarray}
and
\begin{eqnarray}
P^{\bar K} &=& - \frac{1}{2}m_\sigma^2 \sigma^2
+ \frac{1}{2} m_\omega^2 \omega_0^2
+ \frac{1}{2} m_\rho^2 \rho_{03}^2- \frac{1}{2}m_\delta^2 \delta^2+
\Sigma^{0(r)}_{\bar K} \sum_b n^{\bar K}_b \nonumber \\
&& + \frac{1}{3}\sum_b \frac{2J_b+1}{2\pi^2}
\int_0^{k_{F_b}} \frac{k^4 \ dk}{(k^2+m^{* 2}_b)^{1/2}}
+ \frac{1}{3} \sum_l \frac{1}{\pi^2}
\int_0^{K_{F_l}} \frac{k^4 \ dk}{(k^2+m^2_l)^{1/2}} ~,
\end{eqnarray}
where $\Sigma^{0(r)}_{\bar K}$ is the rearrangement term in the antikaon 
condensed phase and has the same form as in Eq. (15), but all quantities in the
equation are replaced by the corresponding quantities of the antikaon condensed
phase. 
Since antikaons form s-wave condensates, they do not contribute to the pressure
directly. Actually the effect of antikaons in the pressure term comes through 
the meson fields.

In the core of neutron stars, various strangeness changing processes 
such as
$N \rightleftharpoons N + \bar K$ and $e^- \rightleftharpoons K^- + \nu_e$ may
occur \cite{Ell,Pal,Bani}.
Here $N\equiv (n,p)$ and $\bar K \equiv (K^-, \bar K^0)$ denote the
isospin doublets for nucleons and antikaons, respectively. 
From the above reactions in chemical equilibrium, we obtain the conditions for 
antikaon condensation \cite{Ell,Pal,Bani}
\begin{eqnarray}
\mu_n - \mu_p &=& \mu_{K^-} = \mu_e ~, \\
\mu_{\bar K^0} &=& 0 ~,
\end{eqnarray}
where $\mu_{K^-}$ and $\mu_{\bar K^0}$ are respectively the chemical
potentials of $K^-$ and $\bar K^0$. 
The charge neutrality condition in the antikaon condensed phase is 
\begin{equation}
Q^{\bar K}=\sum_b q_b n_b^{\bar K} -n_{K^-} -n_e -n_\mu =0.
\end{equation}

It was noted in RMF model calculations that antikaon condensation could be 
either first order or second order phase transition depending on the parameter 
set of the model and antikaon optical potential depth \cite{Sch99,Pal,Bani}. 
If the phase transition is of first order, the mixed phase is 
to be determined by the Gibbs conditions and global
baryon and electric charge conversation laws because we have conserved baryon 
and electric 
charges represented by two chemical potentials $\mu_n$ and $\mu_e$ 
\cite{Gle92}. The Gibbs phase rules read,
\begin{eqnarray}
P^h&=& P^{\bar K},\\
\mu_b^h& =& \mu_b^{\bar K},
\end{eqnarray}
where $\mu_b^h$ and $\mu_b^{\bar K}$ are chemical potentials of baryon b in the
pure hadronic and $K^-$ condensed phase, respectively.
The conditions of global charge neutrality and baryon number conservation are 
imposed through the relations
\begin{equation}
(1-\chi) Q^h + \chi Q^{\bar K} = 0,\\
\end{equation}
\begin{equation}
n_b=(1-\chi) n_b^h + \chi n_b^{\bar K}~,
\end{equation}
where $\chi$ is the volume fraction of $K^-$ condensed phase in the mixed 
phase. The total energy density in the mixed phase is
\begin{equation}
\epsilon=(1-\chi)\epsilon^h + \chi \epsilon^{\bar K}~.
\end{equation}

\section{Parameters}
\subsection{Meson-nucleon couplings}
In DDRH model, the dependence of meson-nucleon vertices on total baryon density
is obtained from microscopic DB calculations of
symmetric and asymmetric nuclear matter. The density dependent vertices in the
RMF model are related to DB self energies in the local density approximation 
\cite{Bro92,Had}. Equating self-energies of infinite nuclear 
matter in RMF and DB calculations, we obtain
\begin{equation}
\Sigma^{RMF}=g_{\alpha} \phi_{\alpha}=\Sigma^{DB},
\end{equation}
where $\phi_\alpha$ represents the field for $\alpha$ meson. Putting the value
of $\phi_{\alpha}$ as given by the meson field equations in the presence of 
nucleons, the above relation simplifies to
\begin{equation}
m^2\Sigma^{DB}=g_{\alpha}^2 n_{\alpha}
\end{equation}
where $n_{\alpha}$ is the density corresponding to $\phi_{\alpha}$ field.
Scalar and vector self energies for neutrons and protons were obtained in DB 
calculations of asymmetric nuclear matter using Groningen potential 
\cite{Len98,Jon98,Mal88}. Using these values of $\Sigma^{DB}$, one immediately 
obtains the density dependent meson-baryon couplings \cite{Hof01}. 
A suitable parameterization for density dependent couplings was made
in Ref.\cite{Hof01}. It has the form 
\begin{equation}
g_{\alpha}(\rho)=a_{\alpha} \frac{1+b_{\alpha}(\rho/\rho_0+d_{\alpha})^2}
{1+c_{\alpha} (\rho /\rho_0 +e_{\alpha})^2}
\end{equation} 
where $\rho_0=0.16 fm^{-3}$ and parameters of the fit are listed in Table I of
Ref.\cite{Hof01}. However, the results of infinite nuclear matter calculation
in DDRH model using the above mentioned parameterization deviated from those of 
DB calculations \cite{Hof01} because momentum dependent DB self energies were
mapped onto the momentum independent DDRH self energies. Therefore, momentum 
dependent vertices with
the additional constraint that the energy density in DB and DDRH are same i.e.
$\varepsilon^{DB}=\varepsilon^{DDRH}$ was proposed \cite{Hof01}. 
Momentum corrected meson-nucleon vertices are given by
\begin{equation}
\tilde g_{\alpha} (k_F) = g_{\alpha} (k_F) \sqrt{1+\zeta_\alpha k_F^2}=\tilde 
\zeta_\alpha (k_F) g_{\alpha}(k_F).
\end{equation}
Momentum corrections $\zeta_\sigma$=0.00804 $fm^2$ and $\zeta_\omega$=0.00103 
$fm^2$ to $\sigma$-nucleon and $\omega$-nucleon were obtained from DB
calculations of symmetric nuclear matter \cite{Hof01}. Using the momentum 
corrected vertices,
Dirac Brueckner EoS for symmetric matter was reproduced well \cite{Hof01}. 
Similarly, momentum correction $\zeta_{\rho}$ to
$\rho$-nucleon vertex was calculated from DB calculations of neutron matter
\cite{Hof01}. This correction was inserted in the DB self energies and the
momentum corrected $\rho$-nucleon vertex was calculated. Later the density 
dependence of the 
momentum corrected $\rho$-nucleon vertex was parameterized using Eq. (42). 
The parameters of this fit are given by Table II in Ref.\cite{Hof01}.
We adopt this parameterization of density dependent couplings and the momentum 
correction prescription in our
calculation. Also, we denote this as Groningen parameter set. In Table I, we 
show meson-baryon couplings for Groningen set at saturation density 
($n_0 = 0.18 fm^{-3}$).
The momentum correction modifies the rearrangement term
as  $\frac {\partial  g_{\alpha B}}{\partial \rho}$ is to be replaced by 
$\frac {\partial \tilde g_{\alpha B}}{\partial \rho}$; this is given by
\cite{Hof01}
\begin{equation}
\frac {\partial \tilde g_{\alpha B} (k_F)}{\partial \rho}=
\tilde \zeta_\alpha (k_F)
\frac {\partial g_{\alpha B} (k_F)}{\partial \rho}
+ \frac {\zeta_\alpha {k_F}
^2 g_{\alpha B} (k_F)}{3 \rho \tilde \zeta_{\alpha B} (k_F)}~.
\end{equation}

In this calculation, we also exploit density dependent meson-nucleon vertices
obtained from DB calculations using Bonn A potential. The parameterization of
vertices is taken from Ref.\cite{Had}. This parameter set is denoted as Bonn A
parameter set. For Bonn A set, $\rho$ meson-nucleon coupling is chosen as 
a constant. Also, $\delta$ meson is not taken into consideration for Bonn A 
potential. Meson-nucleon coupling constants at saturation density 
$n_0 = 0.159 fm^{-3}$ are listed in Table I. 

\subsection{Meson-hyperon couplings}
In the absence of DB calculation including hyperons, density dependence of
meson-hyperon vertices are obtained from density dependent meson-nucleon
couplings using hypernuclei data \cite{Sch96} and scaling law \cite{Kei}. 
This scaling law states that the self energies and vertices of hyperons and 
nucleons are related to each other by their free space coupling constants 
$\bar g_{\sigma Y}$ and $\bar g_{\sigma N}$ \cite{Kei,Hof2},
\begin{equation}
R_{\alpha Y}= \frac{g_{\alpha Y}}{g_{\alpha N}} = \frac{\Sigma_{\alpha Y}}
{\Sigma_{\alpha N}} \approx \frac{\bar g_{\alpha Y}}{\bar g_{\alpha N}}.
\end{equation}
In RMF model, vector meson-hyperon coupling constants were determined from 
scaling factors obtained from SU(6) symmetry relations of the quark model
\cite{Sch96,Dov}. Another possibility is to exploit scaling factors calculated
in microscopic calculations. However, there is only one microscopically derived
free space scaling factor $R_{\sigma \Lambda}$ =0.49 in the literature
\cite{Hai}. We use this value in our calculation. Also, we obtain
the scaling factors for vector and isovector mesons from SU(6) symmetry 
relations \cite{Sch96},
\begin{eqnarray}
\frac{1}{2}g_{\omega \Sigma} = g_{\omega \Xi} =
\frac{1}{3} g_{\omega N},\nonumber\\
\frac{1}{2}g_{\rho \Sigma} = g_{\rho \Xi} = g_{\rho N}{\rm ;}~~~
g_{\rho \Lambda} = 0, \nonumber\\
\frac{1}{2}g_{\delta \Sigma} = g_{\delta \Xi} = g_{\delta N}{\rm ;}~~~
g_{\delta \Lambda} = 0. 
\end{eqnarray}

We obtain $g_{\omega \Lambda}$ 
and scalar meson couplings to other hyperons from the potential depths of 
hyperons in normal nuclear matter. The hyperon potentials in saturated 
nuclear matter are obtained from the experimental data for the single particle
spectra of hypernuclei. In DDRH model, the potential depth of a hyperon (Y) in
saturated nuclear matter is given by 
\begin{eqnarray}
U_Y^N=\Sigma^{0(0)}_Y+\Sigma^{0(r)}_N-\Sigma^{s}_Y~ \label{pot},
\end{eqnarray}
where $\Sigma^{0(0)}_Y = g_{\omega Y} {\omega_0}$, $\Sigma^{s}_Y 
= g_{\sigma Y} {\sigma}$ and $\Sigma^{0(r)}_N$ is the rearrangement 
contribution of nucleons.
In this calculation, the value of $\Lambda$ potential in normal nuclear matter 
is taken as -30 MeV 
\cite{Dov,Chr} and that of $\Xi$ is -18 MeV \cite{Fuk,Kha}. The most updated 
analysis of $\Sigma^-$ atomic data \cite{Bart} and other experimental data
\cite{Fri94} predict a repulsive $\Sigma$-nucleus potential 
depth. Therefore, we adopt a $\Sigma$ well depth of 30 MeV in this calculation 
\cite{Fri94}. We find
that $\Sigma$ hyperons are excluded from the system because of this repulsive
potential. The scaling factors of $\Lambda$ and $\Xi$ 
for Groningen and Bonn A potential are listed in Table II.

From Table II, we observe that $R_{\omega \Lambda}$ is 0.4911 corresponding to 
$R_{\sigma \Lambda} = 0.49$ for Bonn A set. In this case, $R_{\omega \Lambda}$
is obtained from the $\Lambda$ potential depth ($U_{\Lambda}^N$) as discussed 
above. On the other hand, Keil et al. 
\cite{Kei} obtained a value of $R_{\omega \Lambda} = 0.553$ for the same value 
of $R_{\sigma \Lambda}$ from the $\chi^2$ distribution for the deviation of 
DDRH $\Lambda$ single particle energies and hypernuclear data. We perform 
calculations for both the values of $R_{\omega \Lambda}$ for Bonn A set. 

\subsection{Meson-(anti)kaon coupling constants}
Finally, we need to determine the parameter set for meson-(anti)kaon 
interactions.
Here, we do not attribute any density dependence to the vertices of 
meson-(anti)kaons. The vector coupling constants are derived from 
quark model and isospin counting rule so that
\begin{equation}
g_{\omega K} = \frac{1}{3} g_{\omega N} ~~~~~ {\rm and} ~~~~~
g_{\rho K} = g_{\rho N} ~.
\end{equation}
The values of meson-nucleon coupling constants are taken at normal nuclear 
matter density and those are given by Table I.
The scalar coupling constant is obtained from the real part of 
$K^-$ optical potential at normal nuclear matter density
\begin{equation}
U_{\bar K} \left(n_0\right) = - g_{\omega K}\omega_0 - g_{\sigma K}\sigma
+ \Sigma^{0(r)}_N ~.
\end{equation}
The scalar isovector $\delta$ meson also couples with (anti)kaons. The coupling
of $\delta$ meson with (anti)kaons is obtained from simple quark model and this
is given by $g_{\delta K} = g_{\delta N}$. The value of $g_{\delta N}$ is 
obtained from Table I.
 
There are experimental evidences that antikaons experience an attractive 
interaction whereas kaons feel a repulsive interaction in nuclear matter
\cite{Li,Pal2}. It is the depth of antikaon optical potential which is an 
important input in our calculation. The real part of antikaon optical potential
at normal nuclear matter density was evaluated in coupled channel model 
\cite{Koc,Waa} and self consistent calculations \cite{Lut,Ramo,Eff}. These
model calculations give a wide range of values from -120 MeV to -40 MeV for
for $U_{\bar K}$ at $n_0$. Recently, a combined chiral analysis of $K^-$ atomic
and $K^-$p scattering data lead to a shallow attractive $U_{\bar K}(n_0)$ 
of -55 MeV \cite{Cip}. On the other hand, the analysis of $K^-$ atomic data in 
the hybrid model \cite{Fri99} yielded $U_{\bar K} (n_0) = -180\pm20$MeV.
Therefore, there is no consensus among the phenomenological
and microscopic potentials both in terms of depth and $\chi^2$ values from the
fits to kaonic atom data. 
The coupling constants for kaons with $\sigma$-meson, $g_{\sigma K}$,
for a set values of $U_{\bar K}$ from -120 MeV to -180 MeV 
at saturation density for Groningen and Bonn A potential are listed in Table 
III.
The $\sigma$-K coupling constants for Bonn A set are found to be larger than 
that of of Groningen set. This stems mainly from the smaller 
$g_{\omega K} {\omega_0}$ value for Bonn A set compared with that of Groningen
set.   

\section{Results and discussion}

Here we report the results of our calculation in DDRH model using Groningen
set. We perform this calculation for antikaon optical potential 
$U_{\bar K} (n_0)$ = -120 to -180 MeV. There is no $\bar K$ condensation as a 
first 
order phase transition for Groningen set and various values of 
$U_{\bar K}(n_0)$. Rather, $K^-$ and $\bar K^0$ condensation are second order 
phase transitions in this calculation. In this situation, the conditions of 
antikaon condensation are given by Eqs. (32) and (33). Earlier it was found in 
RMF calculations that antikaon condensation could be a second order phase 
transition depending on the antikoan optical potential and coupling constants
of the
models \cite{Sch99,Pal,Bani}. The threshold densities of $K^-$ and $\bar K^0$
condensation in $\beta$-equilibrated matter containing n,p,$\Lambda$ and 
leptons for $U_{\bar K}(n_0)$ = -120 to -180 MeV are recorded in Table IV. 
Besides $\Lambda$ hyperons, we also include other species of hyperons into 
our calculation. However, $\Sigma$ hyperons do not appear because of a repulsive
$\Sigma$-baryon interaction. In Table IV, the critical densities of $\bar K$ 
condensation in $\beta$-equilibrated n,p,$\Lambda$,$\Xi$ and lepton matter are
given in the parentheses. The early appearance of hyperons might have important
effect on the threshold densities of $\bar K$ condensation because hyperons make
the equation of state soft. It was shown in RMF model calculations that the
onset of $\bar K$ condensation was delayed to higher densities due to hyperons
\cite{Ell,Sch96,Kno,Pal,Bani}. Also, negatively charged hyperons diminish the
electron chemical potential delaying the onset of $K^-$ condensation. In this
DDRH model calculation with Groningen set, $\Lambda$ hyperons appear first. 
Consequently, the threshold densities of $\bar K$ condensation are shifted to 
higher densities compared with those in nucleons-only matter.
With further appearance of negatively charged $\Xi^-$ hyperons, $K^-$ 
condensation occurs at higher densities as it is evident from the values in
the parentheses in Table IV. For $U_{\bar K}(n_0)$ = - 120 MeV, 
the early appearance of $\Xi^-$ completely blocks 
the onset of both $K^-$ and $\bar K^0$ condensation 
even in the highest density (8$n_0$) considered in this calculation. On
the other hand, the impact of $\Xi^-$ hyperons on the threshold densities of 
$\bar K^0$ condensation for $|U_{\bar K}(n_0)| \geq 160$ MeV is negligible. 
This may
be attributed to the fact that the density of $\Xi^-$ hyperons falls after the
onset of $K^-$ condensation. This becomes evident when we discuss the
particle density graphs in the following paragraphs. From Table IV, we note 
that the threshold density of $\bar K$ condensation shifts towards lower density
as the strength of $|U_{\bar K}(n_0)|$ increases. This indicates the threshold
of $\bar K$ condensation is not only dependent on the EoS, but also sensitive
to antikaon optical potential depth. For all values of $U_{\bar K}(n_0)$, we 
observe $K^-$ condensation occurs before $\bar K^0$ condensation. 

The composition of neutron star matter containing nucleons (n,p), 
$\Lambda$ hyperon, electron ($e^-$), muon ($\mu^-$) and $K^-$ and $\bar K^0$ 
mesons for Groningen set and $U_{\bar K}(n_0)$ = -160 MeV is
presented in Fig. 1. Before the onset of $K^-$ condensation, the charge 
neutrality is maintained by protons, electrons and muons in the hadronic phase.
We see that $\Lambda$ hyperon is the first strange baryon to appear in the 
hadronic
phase at 1.99$n_0$, where $n_0=0.18 fm^{-3}$. Its density rises fast at the 
cost of neutrons. In this calculation, $K^-$ condensation sets in at 
2.65$n_0$. As soon as $K^-$ condensate appears, it rapidly grows and readily 
replaces $e^-$ and  $\mu^-$. This behaviour is quite expected, as  $K^-$ 
mesons, being bosons, condense in the lowest energy state and are therefore 
energetically favorable to maintain charge neutrality of the system. 
The electron fraction depletes around 4.8$n_0$ 
and the proton density becomes equal to that of $K^-$ condensate. 
The appearance of $\bar K^0$ condensate is delayed
to 6.16$n_0$. With the onset of $\bar K^0$ condensation the abundances of 
n,p,$K^-$ and $\bar K^0$ become identical leading to an isospin 
saturated symmetric matter \cite{Sch99,Pal,Bani}. This may be attributed to
the fact that there is a competition in the production of p-$K^-$ and 
n-$\bar K^0$ pairs resulting in a symmetric matter of nucleons and antikaons
\cite{Sch99,Pal}. Here  
the system is dominated by $\Lambda$ hyperons at high density.
It is worth mentioning here that the results of
DDRH model for Groningen set and $U_{\bar K}(n_0)$ = -160 MeV resemble 
those of the RMF model for GM1 set and $U_{\bar K}(n_0)$ = -160 MeV. However, 
antikaon condensation in the latter case was a first order phase transition. 

Besides $\Lambda$ hyperon formation, we also consider the role of other 
hyperons, such as $\Xi^0$ and $\Xi^-$ on antikaon condensation. In Fig. 2, we 
note that  
negatively charged $\Xi^-$ hyperons start populating the system at 2.24$n_0$, 
soon after the appearance of $\Lambda$. This further postpones $K^-$ 
condensation to 3.20$n_0$. Lepton fractions begin to fall with the 
onset of negatively charged $\Xi^-$. This is quite expected
because it is energetically favourable to achieve charge neutrality among
particles carrying conserved baryon numbers \cite{Gle97}. No such conservation 
law is 
followed by leptons or mesons. But as soon as  $K^-$ condensation sets in, 
lepton fractions as well as $\Xi^-$ fraction drop. It indicates that the EoS
is now mainly softened by the presence of $\Lambda$ hyperons and $K^-$ 
condensate. This has an interesting implication on the threshold density of 
$\bar K^0$ condensation. It is evident from Table IV, $\bar K^0$ 
condensation occurs at 
same density point 6.16 $n_0$ with and without the inclusion of $\Xi^-$ in our 
calculation. 
Also, a symmetric matter of n,p,$K^-$ and $\bar K^0$ condensate emerges here 
after the onset of $\bar K^0$ condensation.
At a much higher density 6.7$n_0$, $\Xi^0$ appears in the system. 

Pressure (P) is plotted against energy density ($\epsilon$)
for various compositions of neutron star matter in Fig. 3. 
Here equations of state (EoS) are calculated with Groningen set. 
The dotted line stands for nucleons-only matter, while the dash-dotted line 
contains $\Lambda$ hyperons in addition to nucleons. The presence of an 
additional degree of freedom softens the EoS appreciably. The solid lines 
correspond to $\Lambda$ hyperon matter including $K^-$ and $\bar K^0$ 
condensate for antikaon optical potentials $U_{\bar K}(n_0)$ = -120 to 
-180 MeV. The kinks on the equations of state mark the onsets of $\bar K$
condensation. 
Already, we have noted that $K^-$ and $\bar K^0$ condensation are second order 
phase transitions for all values of $U_{\bar K}(n_0)$ in our calculation. From
Fig. 3, we find pressure increases with energy density even after the onset of 
antikaon condensation. The appearance
of $K^-$ condensate makes equations of state softer in all cases. The kinks at
higher densities correspond to $\bar K^0$ condensation which further softens 
the EoS. Also, the softness in the EoS is very sensitive to antikaon optical
potential depth. Stronger the attractive antikaon interaction, 
softer is the corresponding EoS. 

The results of static structures of spherically symmetric
neutron stars calculated using Tolman-Oppenheimer-Volkoff (TOV) 
equations \cite{Gle97} and the above mentioned equations of state are now 
presented here. We have used the results of Baym, Pethick and Sutherland 
\cite{Bay} to describe the crust of a compact star
composed  of leptons and nuclei for the low density ($n_B < 0.001 fm^{-3}$)
EoS. In the mid density regime ($0.001 \leq n_B < 0.08 fm^{-3}$)
the results of Negele and Vautherin \cite{Neg} are taken into account. Above 
this density, an EoS calculated in DDRH model has been adopted.
The maximum neutron star masses 
($M_{max}/M_{\odot}$) and their central densities ($u_{cent} = n_{cent}/n_0$) 
for various compositions of matter are listed in Table IV. The values recorded
within the parentheses correspond to the calculations including $\Xi$ hyperons
in addition to $\Lambda$ hyperons. The maximum mass of nucleons-only 
star is 2.313$M_\odot$. The inclusion of $\Lambda$ hyperons softens the 
EoS lowering this value to 1.708$M_\odot$. Because of further softening due to
the inclusion of $\Xi$ hyperons the maximum mass is reduced to a value of
$1.620 M_{\odot}$. The static neutron star sequences representing the stellar
masses $M/M_{\odot}$ and the corresponding central energy densities 
($\varepsilon$) are exhibited in Fig. 4 for n,p,$\Lambda$ and lepton matter 
with $K^-$ and $\bar K^0$ condensate and different values of $U_{\bar K}(n_0)$. 
The softening in the EoS due to the presence of $\bar K$ condensates, leads to 
further lowering in the limiting masses of neutron stars that too attain at 
much earlier central densities as it is evident from Table IV. The maximum mass
of the star varies from 1.697$M_\odot$ (for $U_{\bar K}(n_0)$ = -120 MeV) 
to 1.497$M_\odot$ (for $U_{\bar K}(n_0)$ = -180 MeV) because strong attractive
antikaon interaction in medium produces more softening in the EoS. For 
n,p,$\Lambda$, lepton and $\bar K$ condensate matter composition, 
$K^-$ condensation 
thresholds occur well inside the maximum mass stars for all values of
$U_{\bar K}(n_0)$. So the star is mainly  composed of nucleons, 
$ \Lambda$ hyperons and $K^-$ condensate. 
On the other hand, $\bar K^0$ condensation along with $K^-$ condensation might
be a possibility in maximum mass neutron stars for $|U_{\bar K}(n_0)| \geq 180$ 
MeV. 

We also inspect the effect of $\Xi$ hyperons on the compact star mass 
sequence. Already we have discussed that the appearance of $\Xi^-$ hyperons 
prevents the onset of $\bar K$ condensation for $U_{\bar K}(n_0)$ = -120 MeV.
From Table IV, we find no $\bar K$ condensation occurs inside the limiting
mass neutron stars for $|U_{\bar K}(n_0)| \leq$ 140 MeV. For these values of
$U_{\bar K}(n_0)$, the maximum star mass is the same as that of the case 
without any 
antikaon condensate. On the other hand, $K^-$ condensate is formed inside
maximum mass stars for $|U_{\bar K}(n_0)| \geq$ 160 MeV, but $\bar K^0$ 
condensation
in neutron stars is ruled out for all values of $U_{\bar K}(n_0)$ except for
antikaon potential depth of -180 MeV. For $|U_{\bar K}(n_0)| \geq$ 160 MeV,
we observe there is hardly any change in the maximum masses of neutron stars 
compared with the cases excluding $\Xi$ hyperons. Already we have noted in the
discussion of Fig. 2 that the density of $\Xi^-$ hyperon diminishes with the
appearance of $K^-$ condensate for $|U_{\bar K}(n_0)| \geq$ 160 MeV. Now we
show the equations of state for neutron star matter with and without $\Xi$
hyperons 
for $U_{\bar K}(n_0) =$ -160 MeV in Fig. 5. The solid and dashed line represent
neutron star matter with and without $\Xi$ hyperons respectively. The EoS 
becomes softer in the presence of $\Xi^-$ hyperons, but there is no difference 
between the equations of state just after the onset of $K^-$ condensation. 
This feature is reflected in the maximum masses of neutron stars as is evident
from Table IV. 

In Fig. 6, we draw the mass-radius relationship for n,p,$\Lambda$, lepton matter
with and without $\bar K$ condensate in DDRH model using Groningen set and  
different antikaon optical potential depths and compare 
it with our previous result for hyperonic matter including $K^-$ condensate
calculated in the RMF model \cite{Bani} using GM1 model and 
$U_{\bar K}(n_0)=$ -160 MeV. 
The filled circles correspond to the maximum masses of compact stars.
In case of no $\bar K$ condensate, the maximum mass star has a radius 11.54 km. 
For $U_{\bar K}(n_0) =$ -120 MeV, the maximum mass star 
has a radius 11.76 km, whereas it is 11.39 km for
$U_{\bar K}(n_0) =$ -180 MeV in DDRH model. The smaller radius in the 
latter case may be attributed to more softening in the EoS due to strong 
attractive antikaon potential. The curve corresponding to the RMF calculation 
\cite{Bani} has the smallest radius of 10.9 km among all the cases 
considered here.

We also investigate $K^-$ condensation in nucleons-only and n,p,$\Lambda$ 
matter using Bonn A
set as given by Table I and antikaon optical potential depth at normal
nuclear matter density $U_{\bar K}(n_0) =$ -160 MeV. For Bonn A set, 
$\sigma$-nucleon and 
$\omega$-nucleon couplings are density dependent whereas $\rho$-nucleon 
coupling is a constant one. Here we study the EoS and structure of neutron 
stars. Unlike the situation with Groningen set, antikaon condensation in 
this case is a first order phase transition which is governed by Gibbs phase 
rules and global conservation laws as given by Eqs. (35)-(39). The EoS for
nucleons-only matter with and without $K^-$ condensate are denoted by dotted
and solid line and those of n,p,$\Lambda$ matter are shown by solid and
dashed line in Fig. 7 respectively. For nucleons-only matter, 
antikaon condensation occurs at energy density 397.6 MeV $fm^{-3}$. And 
the phase transition is over at 570.3 MeV $fm^{-3}$. Those two points give 
the extent of the mixed phase. We have pure hadronic phase below the lower 
boundary and antikaon condensed phase above the upper boundary. On the other 
hand, the lower and upper boundary of the first order phase transition to 
$K^-$ condensate in n,p,$\Lambda$ matter for $R_{\sigma \Lambda} = 0.49$ and 
$R_{\omega \Lambda} = 0.4911$ (case I) are shifted to higher energy density
415.0 and 613.2 MeV $fm^{-3}$ respectively, because of the early appearance
of $\Lambda$ hyperons. We find that the effective nucleon mass becomes negative
in n,p,$\Lambda$ matter with and without $K^-$ condensate respectively 
at 6.60$n_0$ and 6.55$n_0$, where $n_0=0.159 fm^{-3}$. We also perform 
calculation for n,p,$\Lambda$ matter with and without
$K^-$ condensate using $R_{\sigma \Lambda} = 0.49$ and 
$R_{\omega \Lambda} = 0.553$ (case II). In this case, $K^-$ condensation 
begins at 397.6 MeV fm$^{-3}$ and the phase transition ends at 
588.4 MeV fm$^{-3}$. Here $\Lambda$ hyperons appear in the mixed phase.
We note the EoS
with and without $K^-$ condensate for case II are stiffer compared with those
of case I. This may be attributed to the stronger repulsion due to the larger 
value of $R_{\omega \Lambda}$ in case II. For case II, we also get negative 
effective nucleon mass in n,p,$\Lambda$ matter with and without $K^-$ 
condensate at 6.72 $n_0$ and 6.60 $n_0$ respectively. It follows from the 
structure 
calculation using TOV equations, the maximum masses of nucleons-only stars 
with and without $K^-$ condensate 
for Bonn A set are 2.55$M_{\odot}$ and 2.32$M_{\odot}$ having central 
densities 5.64 and 6.84 $n_0$, respectively. In this case, we find the radii 
for neutron stars with and without $K^-$ condensate are 10.88 km and 9.91 km, 
respectively. These values of maximum masses and radii in Bonn A set are 
smaller than those of Groningen set. 
For n,p,$\Lambda$ matter with and without $K^-$ condensate  in both case I and 
case II we find that the
effective nucleon mass becomes negative before the maximum masses are reached. 
This feature was earlier found by others for n,p,$\Lambda$ matter without 
antikaon condensate \cite{Hof2}. Because of the behaviour of the
parameterization of the couplings in the high density regime for Bonn A set, the
repulsion due to $\omega$ field becomes larger than the attraction of $\sigma$
field. Consequently, the EoS in Bonn A set is stiffer than that of 
Groningen set. A close inspection of the parameterization of couplings in 
Bonn A set has been already suggested in Ref. \cite{Rin98}. 

\section{Summary and conclusions}
We have studied $K^-$ and $\bar K^0$ condensation in $\beta$-equilibrated
hyperonic matter within a density dependent hadron field 
theoretical model. In this model, baryon-baryon and (anti)kaon-baryon 
interactions are mediated by the exchange of $\sigma$, $\omega$, $\rho$ and
$\delta$ meson. Density dependent 
meson-baryon coupling constants are obtained from microscopic Dirac Brueckner
calculations using Groningen and Bonn A nucleon-nucleon potential. On the other
hand, we have considered constant meson-(anti)kaon couplings in this 
calculation. 

For Groningen set and the values of antikaon optical potential
$U_{\bar K}(n_0) =$ -120 to -180 MeV, both $K^-$ and $\bar K^0$ condensation
are found to be second order phase transition. The early appearance 
of $\Lambda$ hyperons delays $\bar K$ condensation to higher density for all
values of $U_{\bar K}(n_0)$ considered here. With further inclusion of $\Xi$ 
hyperons, $K^-$ as well as $\bar K^0$ condensation do not occur at all for
$|U_{\bar K}|(n_0) <$ 120 MeV, whereas $\bar K$ condensate appears after being 
delayed by $\Xi$ hyperons for $|U_{\bar K}| (n_0) \geq$ 140 MeV. It is 
interesting to
note that as soon as $K^-$ condensate appears in the system, the density of
$\Xi^-$ drops. It is found that antikaon condensation is not only sensitive to 
the equation of state but also to antikaon optical potential depth. 

The equations of state for different neutron star matter compositions including 
$\bar K$ condensate have been studied in DDRH model. 
The appearance of antikaon condensation makes the corresponding EoS softer. 
This softening leads to the reduction in maximum masses of neutron stars for
different cases considered here. For different compositions of neutron star 
matter, it is observed that $K^-$ condensation may occur in maximum mass stars
but the appearance of $\bar K^0$ is ruled out except for $|U_{\bar K}(n_0)| 
\geq 180$ MeV. The neutron star with
smallest maximum mass and radius is obtained for $U_{\bar K}(n_0) =$ -180 MeV 
in DDRH model with Groningen set. We also studied the structure of neutron
stars for nucleons-only matter with and without $K^-$ condensate in DDRH model 
using Bonn A set. In this case, the EoS including $K^-$ condensate
results in a neutron star having radius $<$ 10 km. 

We have compared the results of DDRH model with those of RMF model with GM1 
set \cite{Bani}. The qualitative agreement between the results of these two 
models is
good. Earlier, it was argued that many body correlations may prevent antikaon 
condensation to occur in neutron stars \cite{Pan95,Pan01}. On the contrary, the
study of antikaon condensation in DDRH model with density dependent 
meson-baryon couplings which take into account many body correlations, shows 
that it is a possibility in neutron stars. In this calculation, we have 
treated meson-(anti)kaon couplings as constant. In principle, one may consider
density dependent meson-(anti)kaon couplings. This will introduce additional
rearrangement term in the antikaon sector. It will be reported 
in a future publication.

\newpage

\begin{table} \label{Tab1}
\caption{Density dependent meson-nucleon couplings at saturation density
are obtained from DB calculations using Groningen nucleon-nucleon potential 
in Ref. [34]. 
Infinite nuclear matter properties calculated with momentum
corrected meson-nucleon vertices are binding energy E/A 
= - 15.6 MeV, saturation density $n_0 = 0.18 fm^{-3}$, 
asymmetry energy coefficient $a_{asy} = 26.1$ MeV, incompressibility K = 282
MeV and effective nucleon mass $m^*_N/m_N = 0.554$. Masses for nucleons and 
mesons are $m_N = 939$ MeV, $m_{\sigma} = 550$ MeV, $m_{\omega} = 783$ MeV and
$m_{\rho} = 770$ MeV. The parameterization of density dependent $\sigma$ and
$\omega$-nucleon couplings for Bonn A potential is taken from Ref. [30,31].
The nuclear matter properties in Bonn A potential are   
E/A = - 15.75 MeV, $n_0 = 0.159 fm^{-3}$, 
$a_{asy} = 34.3$ MeV, K = 151.3 MeV and $m^*_N/m_N = 0.642$. All hadronic 
masses for Bonn A case are same as in Groningen case. The $\rho$ meson-nucleon 
coupling is density independent and no $\delta$ meson is present in Bonn A
case. All parameters are dimensionless.}
\hskip 1.5in
\begin{tabular}{|c|c|c|c|c|}
\hfil& $g_{\sigma N}$& $g_{\omega N}$& $g_{\rho N}$&$g_{\delta N}$\\
 \hline
Groningen&  9.9323& 12.1872& 5.6200& 7.6276 \\
Bonn A&9.5105 &11.5401 &8.0758 &- 

\end{tabular}
\end{table}

\hskip 1.5in

\begin{table}\label{tab2}
\caption{Scaling factor for $\sigma$ and $\omega$ meson-hyperon vertices for 
Groningen and Bonn A nucleon-nucleon potential.}
\hskip 1.5in
\begin{tabular}{|c|c|c|c|c|}
\hfil& R$_{\omega \Lambda}$& R$_{\sigma \Lambda}$& R$_{\sigma \Xi}$& R$_{\omega \Xi}$ \\ 
\hline
Groningen& 0.5218&0.49& 0.3104&1/3\\
Bonn A&0.4911& 0.49 &0.3343 &1/3\\
\end{tabular}
\end{table}

\begin{table}\label{tab3}

\caption{The coupling constants for antikaons ($\bar K$) to
$\sigma$-meson, $g_{\sigma K}$, for various values of $\bar K$ optical
potential depths $U_{\bar K}(n_0)$ (in MeV) at saturation density. The
results are for Groningen  and Bonn A nucleon-nucleon potential.}

\hskip 1.5in

\begin{tabular}{|c|c|c|c|c|}

$U_{\bar K}(n_0)$& -120& -140& -160& -180 \\ \hline
Groningen& 0.1993& 0.6738& 1.1483& 1.6228 \\
Bonn A&1.1121 &1.6609 &2.2097 &2.7585

\end{tabular}
\end{table}

\begin{table} 
\caption {The maximum masses $M_{max}$ and their corresponding central
densities $u_{cent}$=$n_{cent}$/$n_{0}$ for nucleon-only (np) star matter and
for stars with
further inclusion of hyperons (np$\Lambda~(\Xi)$) are given below.
The results are for Groningen set.
The threshold densities for $K^{-}$ and $\bar{K}^0$ condensation,
$u_{cr}(K^{-})$ and $u_{cr}(\bar{K}^0)$ where $u = n_B/n_0$ and also 
$M_{max}$ and $u_{cent}$ for neutron star matter including $\Lambda$ hyperons
at different values of antikaon optical potential depth
$U_{\bar{K}}(n_0)$ (in MeV) at saturation density are given. The values 
in the parentheses are those of neutron star matter including $\Xi$.} 
\hskip 1.5in

\begin{tabular}{|c|c|cc|cc|}

{}&${U_{\bar{K}}(n_0)}$&${u_{cr}(K^{-})}$&${u_{cr}(\bar{K}^{0})}$&${u_{cent}}$
&${M_{max}/M_{\odot}}$\\ \hline
&&&&&\\
{np}&{-}&{-}&{-}&{5.11}&{2.313}\\ \hline
&&&&&\\
{np$\Lambda(\Xi)$}&{}&{-}&{-}&{5.13~(4.89)}&{1.708~(1.620)}\\ \hline
&&&&&\\
{}&{-120}&{3.83~(-)}&{-~(-)}&{4.84~(4.89)}&{1.697~(1.620)}\\
{np$\bar K \Lambda (\Xi)$}&{-140}&{3.17~(5.74)}&{7.27~(7.39)}&{4.56~(4.89)}&{1.665~(1.620)}\\
{}&{-160}&{2.65~(3.20)}&{6.16~(6.16)}&{4.38~(4.49)}&{1.602~(1.599)}\\
{}&{-180}&{2.28~(2.29)}&{5.16~(5.16)}&{5.16~(5.16)}&{1.497~(1.497)}\\ 
\end{tabular}
\end{table}
\newpage 

{\large{\bf Figure Captions}}
\vspace{0.5cm}

FIG. 1. Number densities ($n_i$) of various particles in 
$\beta$-equilibrated n,p,$\Lambda$ and lepton matter including $K^-$ and 
$\bar K^0$ condensate for Groningen set
and  antikaon optical potential depth at normal nuclear matter density 
$U_{\bar K}(n_0) = -160$ MeV as a function of normalised baryon density.
\vspace{0.5cm}

FIG. 2. Number densities ($n_i$) of various particles in 
$\beta$-equilibrated n,p,$\Lambda$,$\Xi$ and lepton matter including $K^-$ and 
$\bar K^0$ condensate for Groningen set
and  antikaon optical potential depth at normal nuclear matter density 
$U_{\bar K}(n_0) = -160$ MeV as a function of normalised baryon density.
\vspace{0.5cm}

FIG. 3. The equation of state, pressure $P$ vs. energy density $\varepsilon$ 
for Groningen set is shown here. The results are for  n,p and lepton matter 
(dotted line), n,p,$\Lambda$ and lepton matter (dash-dotted line), 
and n,p,$\Lambda$ and 
lepton matter including $K^-$ and $\bar K^0$ condensate (solid lines)
calculated with antikaon optical potential depth at normal nuclear matter 
density of $U_{\bar K}(n_0) = -120, -140, -160, -180$ MeV. 
\vspace{0.5cm}

FIG. 4. The compact star mass sequences are plotted with central energy 
density for Groningen set and antikaon optical potential depth 
of  $U_{\bar K}(n_0)=-120, -140, -160, -180$ MeV. The star masses of 
n,p,$\Lambda$ and lepton matter with $K^-$ and $\bar K^0$ condensate are shown  
here. 
\vspace{0.5cm}

FIG. 5. The equation of state for 
n,p,$\Lambda$,$\Xi$, lepton and $\bar K$ matter (solid line) and n,p,$\Lambda$,
lepton and $\bar K$ matter (dashed lines) calculated with Groningen set and 
antikaon optical potential depth at normal nuclear matter 
density of $U_{\bar K}(n_0) = -160$ MeV are shown. 
\vspace{0.5cm}

FIG. 6. The mass-radius relationship for compact star sequences for n,p,
$\Lambda$ and lepton matter with $K^-$ and $\bar K^0$ 
condensate for Groningen set and antikaon optical potential depth of 
$U_{\bar K}(n_0) = -120, -140, -160, -180$ MeV. The mass-radius relationship 
for compact star sequence for hyperon matter including $K^-$ 
condensate in the RMF model calculation (Ref. [9]) is also shown here.
\vspace{0.5cm}

FIG. 7. The equation of state for n,p and lepton matter (solid line) 
and n,p, lepton and $K^-$ matter (dotted line) calculated with Groningen 
set and antikaon optical potential depth at normal nuclear matter density of 
$U_{\bar K}(n_0) = -160$ MeV are shown. The equation of state for 
n,p,$\Lambda$ matter with and without $K^-$ condensate for different values 
of $R_{\omega \Lambda}$ are also plotted here. 

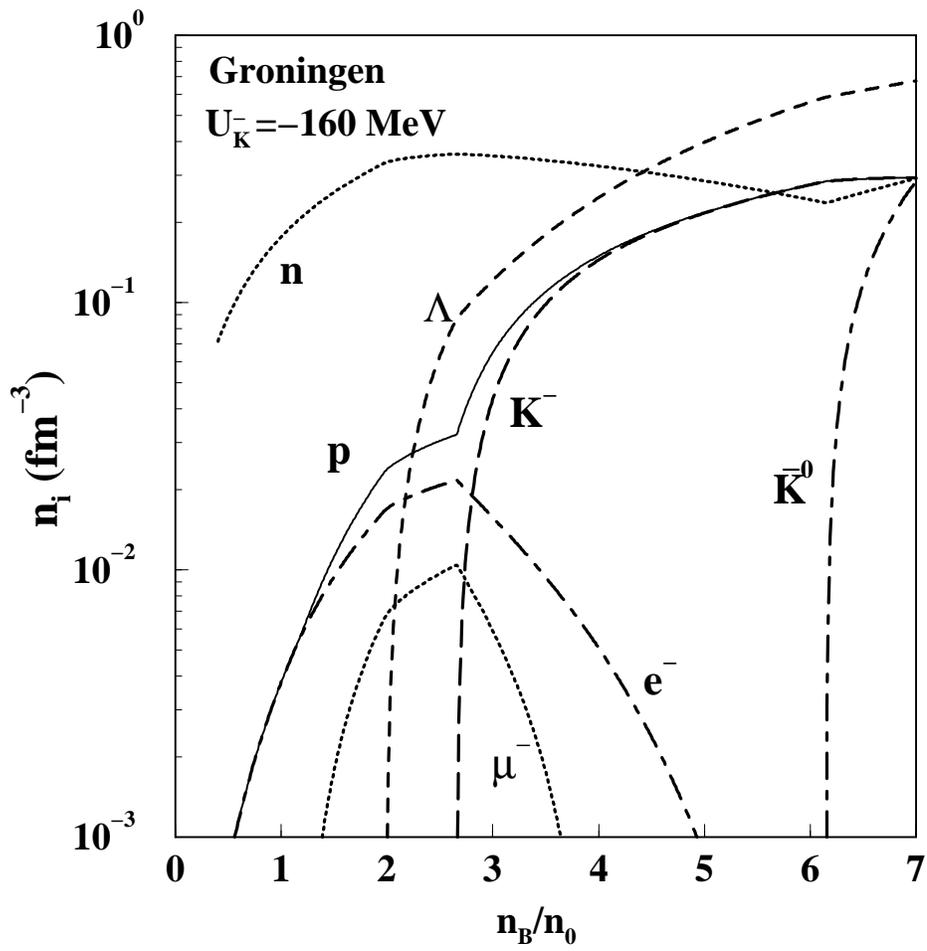
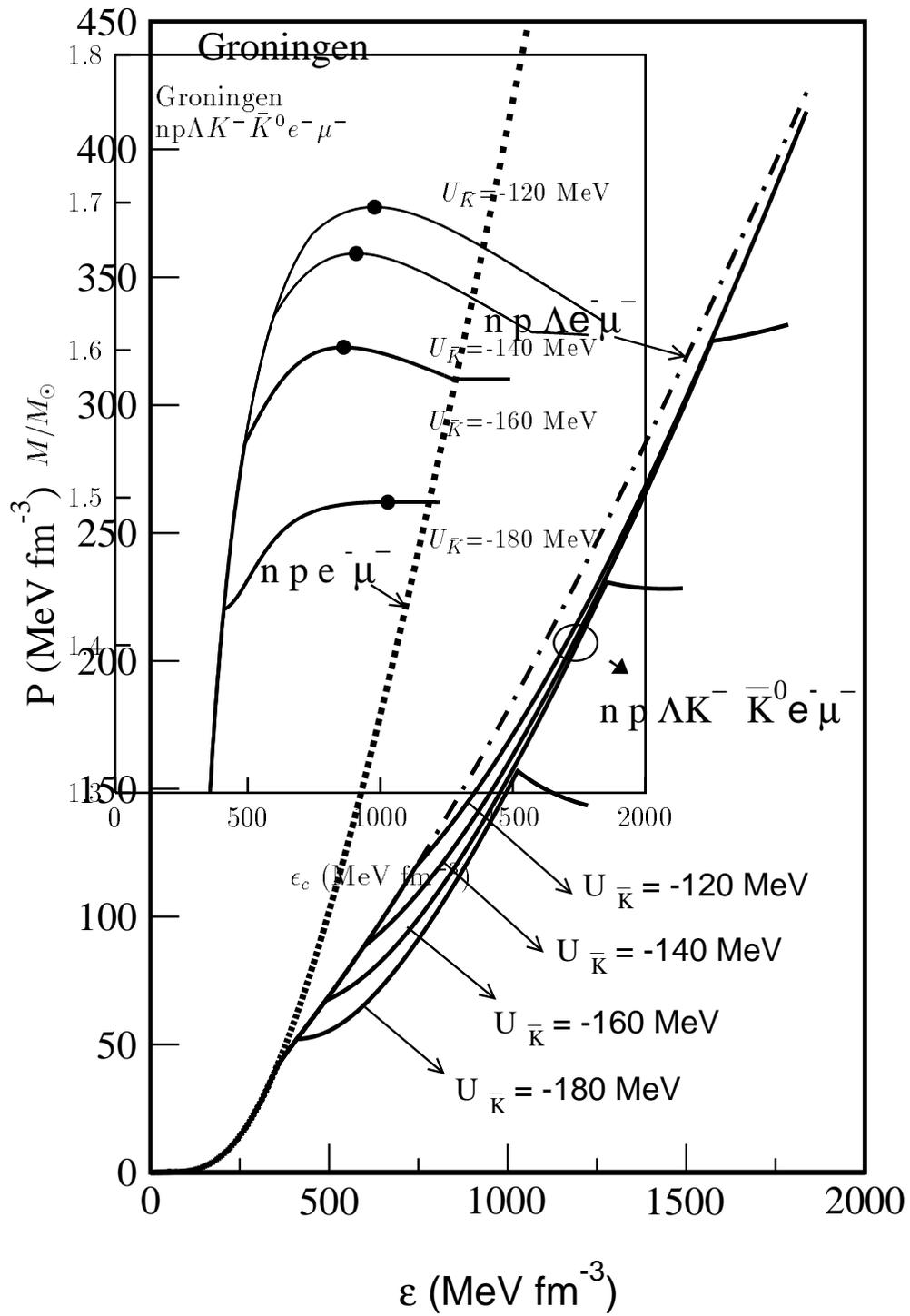
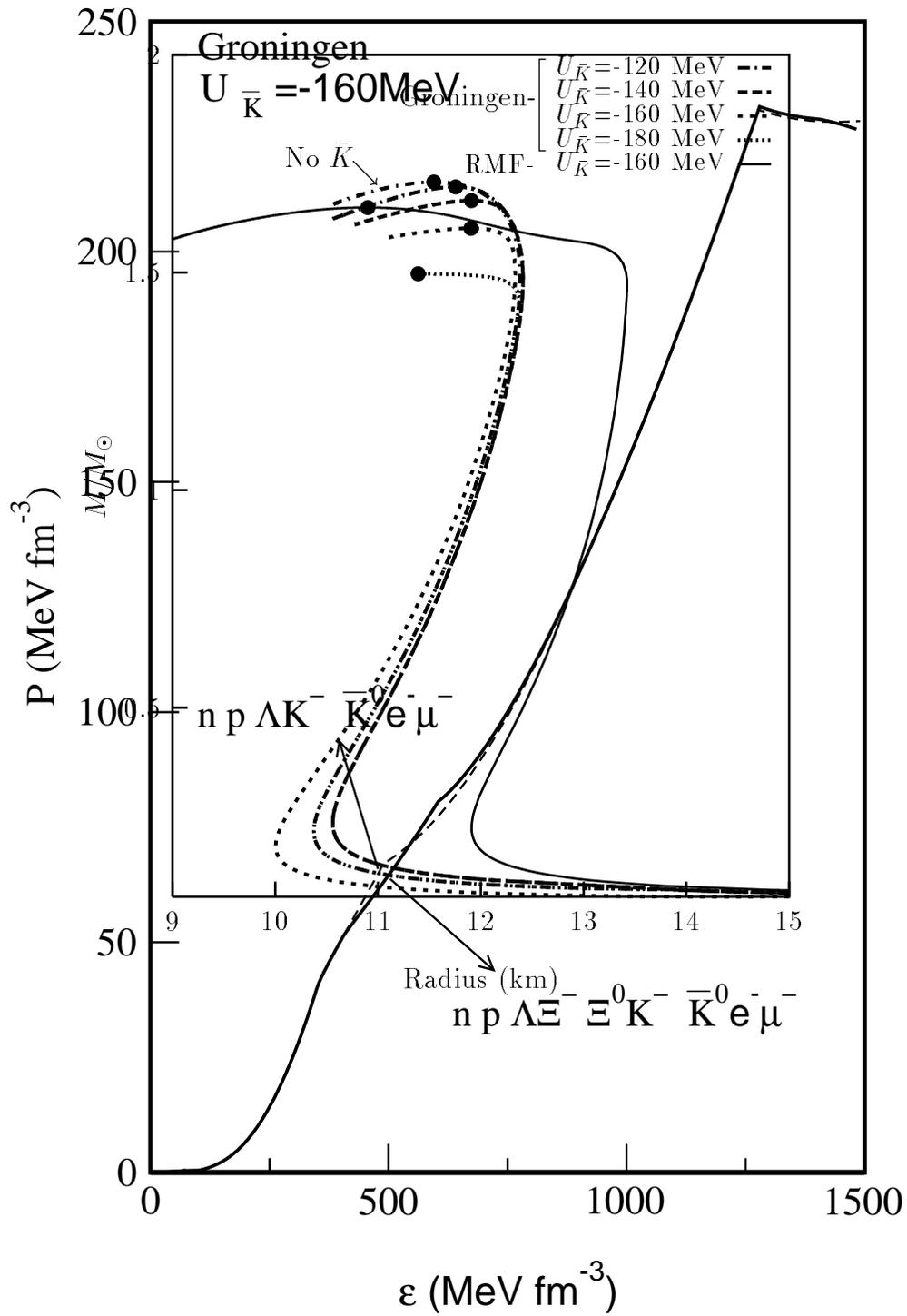
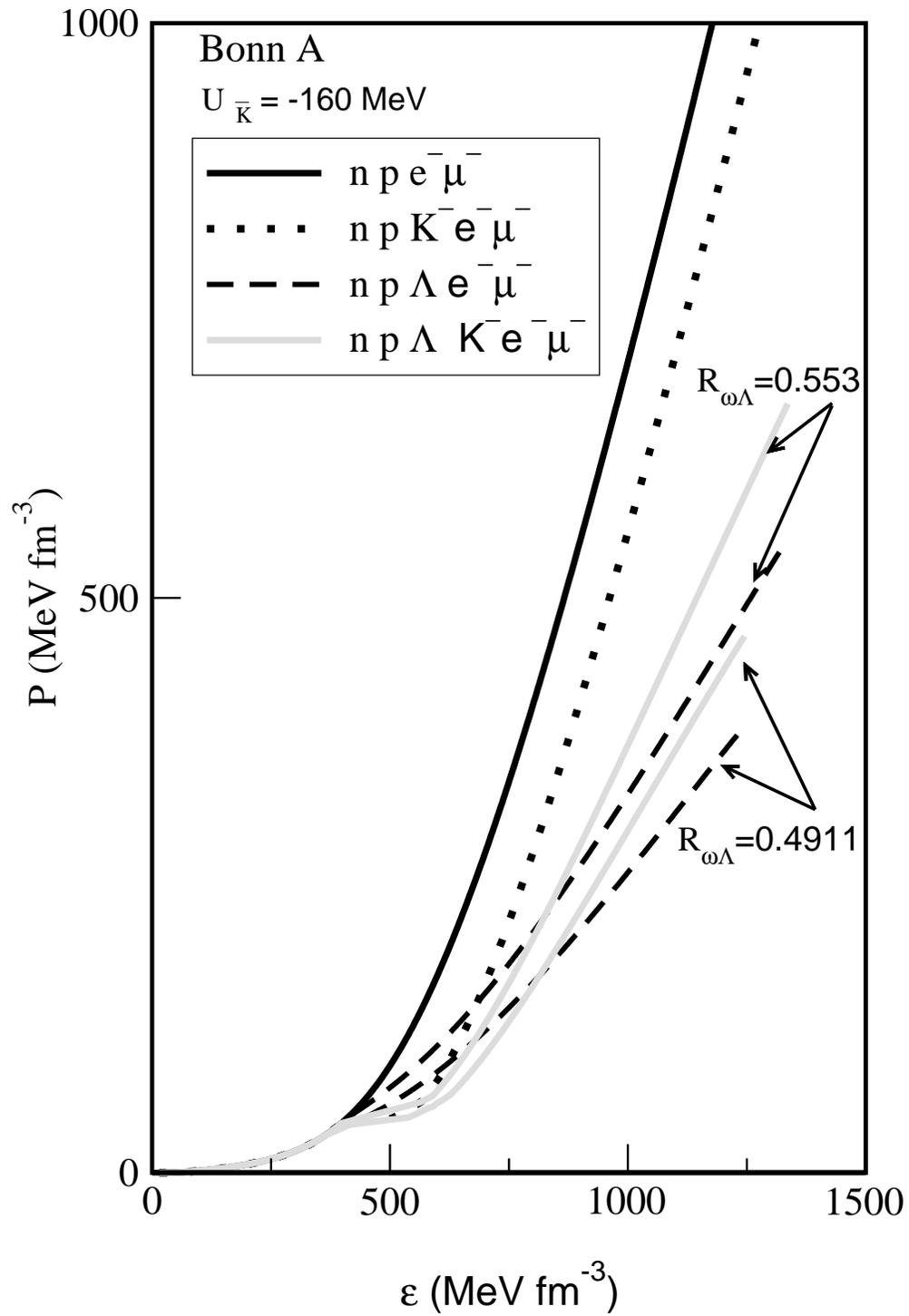
\end{document}